\documentclass[journal]{IEEEtran}
\usepackage{spconf,amsmath}
\usepackage{amsthm}
\usepackage{amsfonts}
\usepackage{amssymb}
\usepackage{mathtools}
\usepackage{float}
\usepackage{algorithm,algorithmic}
\usepackage[pdftex]{graphicx}
\usepackage{amsmath}
\usepackage{color}
\usepackage{multirow}
\usepackage{graphicx}
\usepackage{bm}
\usepackage{balance}
\usepackage{comment}
\usepackage{cite}
\usepackage[caption=false,font=footnotesize]{subfig}

\newtheorem{theorem}{Theorem}%[section]

\newtheorem{proposition}[theorem]{Proposition}

\title{Beurling-Selberg Extremization for Dual-Blind Deconvolution Recovery in Joint Radar-Communications}

\name{Jonathan Monsalve$^{ *}$, Edwin Vargas$^{\dagger}$, Kumar Vijay Mishra$^{\ddagger}$, Brian M. Sadler$^{\ddagger}$ and Henry Arguello$^{\dagger}$}
\address{$^{\dagger}$Universidad Industrial de Santander, Bucaramanga, Colombia, 680002 \\
$^{*}$Corporación Universitaria Minuto de Dios, Bogotá, Colombia, 111021 
\\ $^{\ddagger}$United States DEVCOM Army Research Laboratory, Adelphi, MD 20783 USA
\thanks{This research was sponsored by the Army Research Office/Laboratory under Grant Number W911NF-21-1-0099, and the VIE project entitled ``Dual blind deconvolution for joint radar-communications processing''. K. V. M. acknowledges support from the National Academies of Sciences, Engineering, and Medicine via the Army Research Laboratory Harry Diamond Distinguished Fellowship.}}

\ninept

\begin{document}

\maketitle
\IEEEpeerreviewmaketitle

\begin{abstract}
Recent interest in integrated sensing and communications has led to the design of novel signal processing techniques to recover information from an overlaid radar-communications signal. Here, we focus on a spectral coexistence scenario, wherein the channels and transmit signals of both radar and communications systems are unknown to the common receiver. In this dual-blind deconvolution (DBD) problem, the receiver admits a multi-carrier wireless communications signal that is overlaid with the radar signal reflected off multiple targets. The communications and radar channels are represented by continuous-valued range-times or delays corresponding to multiple transmission paths and targets, respectively. Prior works addressed recovery of unknown channels and signals in this ill-posed DBD problem through atomic norm minimization but contingent on \textit{individual} minimum separation conditions for radar and communications channels. In this paper, we provide an optimal \textit{joint} separation condition using extremal functions from the Beurling-Selberg interpolation theory. Thereafter, we formulate DBD as a low-rank modified Hankel matrix retrieval and solve it via nuclear norm minimization. We estimate the unknown target and communications parameters from the recovered low-rank matrix using \textit{mu}ltiple \textit{si}gnal \textit{c}lassification (MUSIC) method. We show that the joint separation condition also guarantees that the underlying Vandermonde matrix for MUSIC is well-conditioned. Numerical experiments validate our theoretical findings.
\end{abstract}
\begin{IEEEkeywords}
Beurling-Selberg majorant, condition number, dual-blind deconvolution, joint radar-communications, passive radar.
\end{IEEEkeywords}

\section{Introduction}
\label{sec:intro}
The electromagnetic spectrum is a scarce natural resource. With the advent of cellular communications and novel radar applications, the spectrum has become increasingly contested. This has led to the development of joint radar-communications (JRC) systems, which facilitate spectrum-sharing while offering benefits of low cost, compact size, and less power consumption \cite{mishra2019toward,duggal2020doppler,elbir2021terahertz,elbir2022the}. The JRC design topologies broadly fall into three categories: co-design \cite{liu2020co}, cooperation \cite{bicua2018radar}, and co-existence \cite{wu2022resource}. The spectral co-design employs a common transmit waveform and/or hardware while the cooperation technique relies on opportunistic processing of signals from one system to aid the other. The spectral co-existence is useful for legacy systems, wherein the radar and communications transmit and access the channel independently, receive overlaid signals at the receiver, and mitigate mutual interference. In this paper, we focus on the overlaid receiver for the spectral coexistence scenario.

In conventional radar applications, the transmit waveform is known to the receiver and the goal of signal processing is to extract the unknown target parameters. In wireless communications, the roles are reversed with the channel estimated \textit{a priori} and the receiver estimating the unknown transmit messages. In certain applications, both signals and channels may be unknown to the receiver. For instance, passive \cite{sedighi2021localization} and multistatic \cite{dokhanchi2019mmwave} radars employed for low-cost and efficient covert operations are generally not aware of the transmit waveform \cite{kuschel2019tutorial}. In mobile radio \cite{neskovic2000modern} and vehicular \cite{olariu2009vehicular} communications, the channel is highly dynamic and any prior estimates may be inaccurate. A common receiver \cite{vouras2022overview} in this general spectral coexistence scenario, therefore, deals with unknown radar and communications channels and their respective unknown transmit signals. 

Prior works modeled the extraction of all four of these quantities, i.e. radar and communications channels and signals, as a \textit{dual-blind deconvolution} (DBD) problem\cite{vargas2023dual,vargas2022joint,jacome2022multid,jacome2023factor}. Herein, the observation is a sum of two convolutions and all four signals being convolved need to be estimated. This formulation is related to (single-)blind deconvolution (BD), a longstanding problem that occurs in a variety of engineering and scientific applications \cite{jefferies1993restoration,ayers1988iterative,abed1997blind}. The DBD problem is highly ill-posed. In \cite{vargas2022joint}, authors leveraged upon the sparsity of radar and communications channels to recast DBD as the minimization of the sum of multivariate atomic norms (SoMAN). The ANM facilitates recovering continuous-valued parameters \cite{bhaskar2013atomic,mishra2015spectral,xu2014precise}. Using the theories of positive hyperoctant polynomials, they then devised a semidefinite program (SDP) for SoMAN and estimated the unknown target and communications parameters.  

The SoMAN approach guarantees perfect DBD recovery assuming individual minimum separation conditions of the spikes (or targets) in the radar and communications channels \cite{vargas2022joint}. In this paper, we provide an improved guarantee by suggesting a joint minimum separation. In particular, we focus on delay-only DBD and rely on relating the extremal functions in the Beurling-Selberg interpolation theory \cite{graham1981class,10.2307/23513054, 10.2307/25747145, Carneiro2011} to the spectral properties of an entry-wise weighted Vandermonde matrix that results from the decomposition of the vectorized Hankel matrix. We employ the Beurling-Selberg majorant and minorant functions of the interval function on the real line because they are compactly supported on the Fourier domain. The connection between extremal functions and the condition number of Vandermonde matrices was earlier studied by \cite{moitra2015super} in the context of (non-blind) super-resolution.

Contrary to the SoMAN formulation that exploits channel sparsity, our approach is based on nuclear norm minimization to recover a low-rank vectorized Hankel matrix, from which unknown target and communications parameters are estimated by the \textit{mu}ltiple \textit{si}gnal \textit{c}lassification (MUSIC) algorithm \cite{chen2022vectorized}. The low-rankness of radar/communications receive signals has been previously validated in related applications such as super-resolution of complex exponentials \cite{yang2016super} and multipath channels in wireless communications \cite{ahmed2013blind,luo2006low}. In this work, our joint separation condition guarantees a well-conditioned Vandermonde matrix -- shown to be low-rank in a sliding column form \cite{heckel2016super} -- for MUSIC-based recovery.  

This paper uses boldface lowercase and uppercase for vectors and matrices, respectively. The $i$-th entry of the vector $\mathbf{x}$ is $[{\mathbf{x}}]_i$, $[{\mathbf{X}}]_{i,j}$ represents the $i$-th row and $j$-th column of $\mathbf{X}$. ${\mathbf{x_r}}_{i}$ the $i$-th column of $\mathbf{X_r}$. We denote the transpose, conjugate, and Hermitian by $(\cdot)^T$, $(\cdot)^*$, and $(\cdot)^H$, respectively. Unless otherwise specified, the integrals and summation limits are $-\infty$ and $\infty$.  The functions $\text{max} (a,b)$ and $\text{min}(a,b)$ return, respectively, maximum and minimum of the input arguments. The expression $\sum_{i,i'=1}^{K,Q} a_i b_{i'} = \sum_{i=1}^{K} \sum_{i'=1}^Q a_i b_{i'}$. $\mathcal{F}$ represents the continuous-time Fourier transform (CTFT); $\ast$ denotes the convolution operation; $\langle \mathbf{A},\mathbf{B} \rangle = \textrm{Tr}( \mathbf{A}^*\mathbf{B})$, and $\textrm{Tr}(\cdot)$ is the matrix trace;  $\sigma_{\textrm{min}}(\mathbf{A})$ is the minimum singular value of $\mathbf{A}$; and rank($\mathbf{A}$) is its matrix rank.
\vspace{-8pt}
\section{Signal Model}
\label{sec:sysmod}
Consider a radar that transmits a time-limited baseband pulse $g_r(t)$, whose CTFT is $\widetilde{g}_r(f) = \int g_r(t)e^{-j2\pi ft}\partial t $, assuming that most of the radar signal's energy lies within the frequencies $\pm B/2$ yields to $g_r(f) \approx \int_{-B/2}^{B/2} g_r(t)e^{-j2\pi ft}\partial t$. The pulse $g_r(t)$ is reflected back to the receiver by $K$ targets, where the $i$-th target is characterized by time delay $[\bm{\tau_r}]_i$ and complex amplitude $[\bm{\beta}]_i$. $[\bm{\tau_r}]_i$ is linearly proportional to the target's range and $[\bm{\beta}]_i$ models the path loss and reflectivity. The radar channel is
\begin{equation}
    h_r(t) = \sum_{i=0}^{K-1} [\bm{\beta}]_i \delta(t-[\bm{\tau_r}]_i).
    \label{eq:hr}
\end{equation}
%
%where $[\bm{\tau_r}]_i \in [0,1) $ is the delay associated with the $i$-th target range, $\bm{\tau_r} \in \mathbb{R}^K$, and $\bm{\beta}$ is the vector with complex amplitudes. 

The communications transmitted signal $g_c(t)$ is a message modeled as an orthogonal frequency-division multiplexing (OFDM) signal with bandwidth $B$ and $M$ equi-bandwidth sub-carriers separated by $\Delta f$, i.e.,
\begin{equation}
    g_c(t) = \sum_{l=1}^{M} [\widetilde{\mathbf{g}}_c]_l e^{\textrm{j}2\pi l \Delta f t},
\end{equation}
where $[\widetilde{\mathbf{g}}_c]_l$ is the modulated symbol onto the $l$-th subcarrier. The message propagates through a communications channel $h_c(t)$ that comprises $Q$ paths characterized by their attenuation/channel coefficients $\bm{\omega} \in \mathbb{C}^{Q}$ and delays $\bm{\tau_c} \in \mathbb{C}^{Q}$: %. Thus, the communications channel is modeled as 
\begin{equation}
    h_c(t) = \sum_{q=1}^{Q} [\bm{\omega}]_q \delta(t-[\bm{\tau_c}]_q).
    \label{eq:hc}
\end{equation}

The radar and communications systems share the spectrum in a spectral coexistence scenario. The common received signal is a superposition of radar and communications signals propagated through their respective channels as
\begin{align}
    y(t) &= g_r(t) \ast h_r(t) + g_c(t) \ast h_c(t), \label{eq: join}\nonumber \\
    &= \sum_{i=1}^K [\bm{\beta}]_i g_r(t-[\bm{\tau_r}]_i) + \sum_{q=1}^Q [\bm{\omega}]_q g_c(t-[\bm{\tau_c}]_q).
\end{align}
The CTFT of the overlaid signal is 
\begin{align}
 Y(f) =\sum_{i=1}^{K} [\bm{\beta}]_i e^{-\textrm{j}2\pi [\bm{\tau}_r]_i f}  \widetilde{g}_r(f) + \sum_{q=1}^{Q} [\bm{\omega}]_q e^{-\textrm{j}2\pi [\bm{\tau}_c]_q f} \widetilde{g}_c(f),
 \label{eq:f1}
\end{align}
where $\widetilde{g}_c(f) = \mathcal{F}\{g_c(t)\}$. Sampling \eqref{eq:f1} at the Nyquist rate of $\Delta_f = \frac{B}{N}$, where $n=0,\ldots,N-1$, yields
\small
\begin{align}
 Y(n) =
 &\sum_{i=1}^{K} [\bm{\beta}]_i e^{-\textrm{j}2\pi [\bm{\tau}_r]_i n\Delta_f}  \widetilde{g}_r(n\Delta_f)\nonumber \\ &+ \sum_{q=1}^{Q} [\bm{\omega}]_q  e^{-\textrm{j}2\pi [\bm{\tau}_c]_q n\Delta_f} \widetilde{g}_c(n\Delta_f).
\end{align}
Collect the samples of $\widetilde{g}_r(n\Delta_f)$ in the vector $\widetilde{\mathbf{g}_r}$, i.e, $[\widetilde{\mathbf{g_r}}]_n=\widetilde{g}_r(n\Delta_f)$. Similarly, $\widetilde{g}_c (n\Delta_f) = [\widetilde{\mathbf{g_c}}]_n$. The samples $[\mathbf{y}]_n=Y(n)$ of the observation vector $\mathbf{y}$ are
\begin{align}
    [\mathbf{y}]_n &= \sum_{i=1}^{K} [\bm{\beta}]_i e^{-\textrm{j} 2\pi [\bm{\tau}_r]_i n}  [\widetilde{\mathbf{g_r}}]_n + \sum_{q=1}^{Q} [\bm{\omega}]_q e^{-\textrm{j}2\pi [\bm{\tau}_c]_i n} [\widetilde{\mathbf{g_c}}]_n.
    \label{eq:problem1}
\end{align}
Our goal in the DBD problem is to estimate the parameters $\bm{\tau}_r$, $\bm{\tau}_c$, $\bm{\beta}$, and $\bm{\omega}$ from $\mathbf{y}$, without knowledge of the radar pulse $\widetilde{\mathbf{g}}_r$ or communications symbols $\widetilde{\mathbf{g}}_c$. This is an ill-posed problem because of many unknown variables and fewer measurements. Finally, the problem can be expressed in matrix form by denoting $\alpha_{[\bm{\tau}_r]_{i}}^n = e^{-\textrm{j}2\pi n [\bm{\tau}_r]_{i}}$, $\alpha_{[\bm{\tau}_c]_{l}}^n = e^{-\textrm{j}2\pi n [\bm{\tau}_c]_{l}}$, and $\bm{\psi}=[\bm{\beta}^T,\bm{\omega}^T]^T$. Rewrite the overlaid receiver signal in \eqref{eq:problem1} as a linear system:
    \begin{align}
        &\mathbf{y} 
            =\begin{bsmallmatrix}
                [\widetilde{\mathbf{g_r}}]_0 & . & [\widetilde{\mathbf{g_r}}]_1 & [\widetilde{\mathbf{g_c}}]_1 & . & [\widetilde{\mathbf{g_c}}]_1\\
                [\widetilde{\mathbf{g_r}}]_2\alpha_{[\bm{\tau}_r]_1}^1 & . & [\widetilde{\mathbf{g_r}}]_2 \alpha_{[\bm{\tau}_r]_{K}}^1 & [\widetilde{\mathbf{g_c}}]_2\alpha_{[\bm{\tau}_c]_1}^1 & . & [\widetilde{\mathbf{g_c}}]_2\alpha_{[\bm{\tau}_c]_{Q}}^1\\
                : & . & : & :  & . & :  \\
              [\widetilde{\mathbf{g_r}}]_{N}\alpha_{[\bm{\tau}_r]_1}^{N} & . & [\widetilde{\mathbf{g_r}}]_{N}\alpha_{[\bm{\tau}_r]_{K}}^{N} & [\widetilde{\mathbf{g_c}}]_{N}\alpha_{[\bm{\tau}_c]_1}^{N} & . & [\widetilde{\mathbf{g_c}}]_{N}\alpha_{[\bm{\tau}_c]_{Q}}^{N}
            \end{bsmallmatrix}\bm{\psi}\nonumber\\
            &= \text{diag}(\mathbf{g_r})\mathbf{V_r}\bm{\beta} +  \text{diag}(\mathbf{g_c})\mathbf{V_c}\bm{\omega}= \mathbf{V_g} \bm{\psi}\label{eq:problemjoint},
    \end{align}
where $\mathbf{V_r}$ and $\mathbf{V_c}$ are Vandermonde matrices such that  $[\mathbf{V_r}]_{\iota,i}=\alpha_{[\bm{\tau}_r]_{i}}^\iota$ and $[\mathbf{V_c}]_{\iota,l}=\alpha_{[\bm{\tau}_c]_{l}}^\iota$.
\vspace{-8pt}
\section{Beurling-Selberg Extremal functions}% Analysis for DBD}
The guarantees to DBD recovery in previous studies \cite{vargas2023dual} are based on the minimum separation of spikes in each channel assuming a noiseless scenario. Here, we resort to the Beurling-Selberg extremal functions to derive an optimal separation condition that imposes a joint structure on the two channels. This result follows analyzing the condition number of $\mathbf{V_g}$ that controls the stability of the solution of \eqref{eq:problemjoint} in the presence of additive noise. %The stability of the solution of \eqref{eq:problemjoint} in the presence of additive noise depends on the condition number of $\mathbf{V_g}$. 
%--------------------------------------------------------------
\begin{figure}[t]
    \centering
    \includegraphics[width=0.45\textwidth]{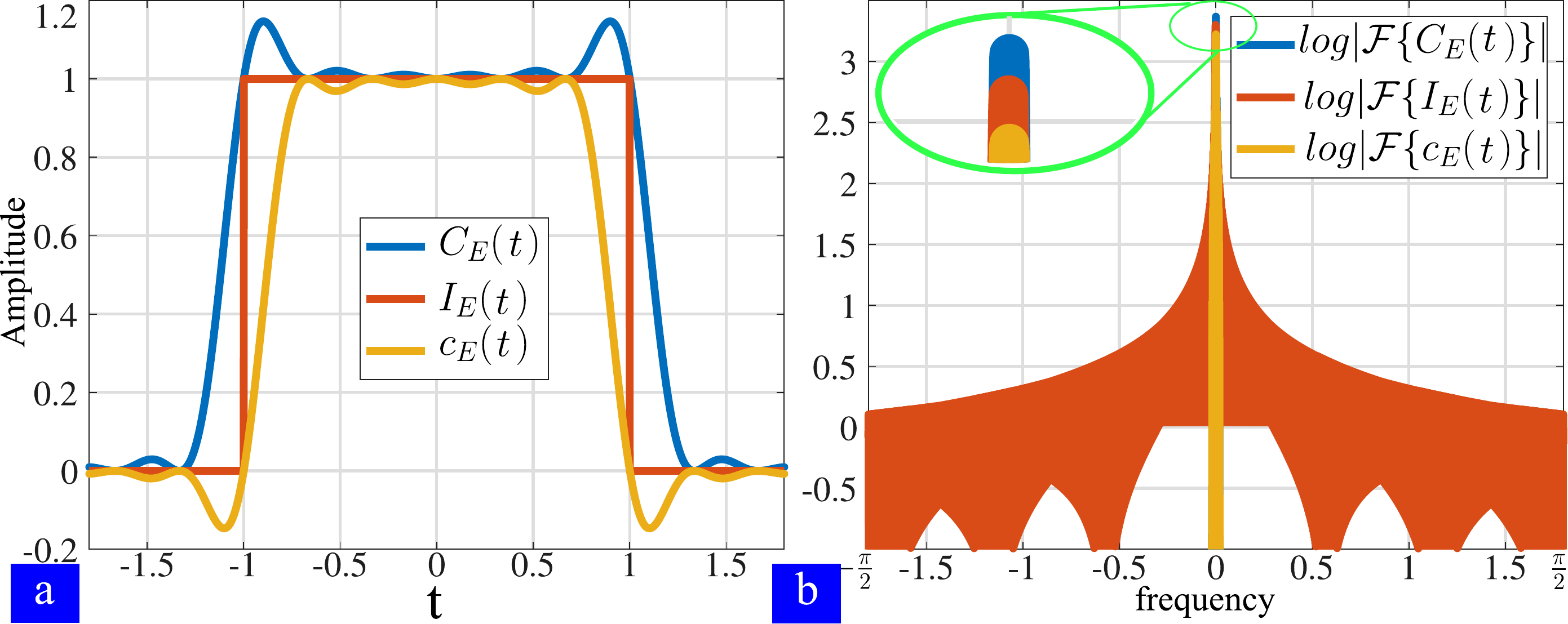}
    \caption{(a) Beurling-Selberg majorant (blue) and minorant (orange) functions and the indicator function $I_E(t)$ (red) for the interval $E=[-1,1]$. (b) Fourier transform of $C_E(t)$ (blue), $I_E(t)$ (red), and $c_E(t)$ (orange) on log-absolute-value scale. }
    \label{fig:beurlingselberg}
\end{figure}
%--------------------------------------------------------------

%that is, functions whose growth is bounded by $Ce^{(2\pi + \epsilon) |z|}$ for every $\epsilon>0$, $z \in \mathbb{C}$, and constant $C$

%The conditioning of this matrix is analyzed via \textcolor{blue}{extremal functions} \textcolor{red}{undefined term} in the Beurling-Selberg interpolation theory. 

%Beurling \cite{10.2307/25747145, collectedB} found a solution to the \textcolor{blue}{extremal problem} \textcolor{red}{undefined term}  of finding real entire functions $b(t)$ and $B(t)$ of type $2\pi$, that satisfy $b(t) \leq \text{sign}(t) \leq B(t)$ and minimize the integrals $\int B(t) - \text{sign}(t) \partial t$ %,  and $\int  \text{sign}(t) - b(t) \partial t$. 
Consider the indicator function $I_E(t)$ defined on the interval $[0,N-1]$, where $I_E(t)$ equals $1$ if $0 < t < N-1$, and $0$ otherwise. We can find approximations for $I_E(t)$ using functions $b(t)$ and $B(t)$ such that $b(t) \leq I_E(t) \leq B(t)$. Moreover, we seek to minimize the integrals $\int (B(t) - I_E(t)) dt$ and $\int (I_E(t) - b(t)) dt$. These optimization problems are known as extremal problems, and the resulting functions $C_E(t)$ (majorant) and $c_E(t)$ (minorant) are considered extremal functions.

Beurling \cite{10.2307/25747145, collectedB} found a solution to this extremal problem for the sign function by restricting $C_E(t)$ and $c_E(t)$ to be real entire functions of exponential type $2\pi$. Recall that if an entire function $F: C^N \rightarrow C $ is real-valued when it is restricted to $R^N$, it is called a real entire function. Additionally, if this statement holds and the growth of $F$ is bounded by a constant $C$ and $Ce^{(2\pi + \epsilon) |z|}$ for every $\epsilon>0$, $z \in \mathbb{C}$, it is called a real entire function of exponential type $2\pi$. Selberg observed that Beurling functions can majorize and minorize the characteristic function $I_E(t)$. The resultant majorant and minorant functions have the useful property that their Fourier transforms are continuous functions supported on the interval $[-\Delta,\Delta]$ \cite{graham1981class,10.2307/23513054, 10.2307/25747145, Carneiro2011}. We intend to employ these functions to provide a bound on the $\mathbf{V_g}$ condition number. Fig. \ref{fig:beurlingselberg}a shows that these functions approximate and bound the indicator function as $C_E(t)\leq I_E(t)\leq c_E(t)$. Fig. \ref{fig:beurlingselberg}b depicts these functions in the Fourier domain  To this end, define the separation between the delay parameters corresponding to radar-only, communications-only and radar-communications measurements as, respectively, $\gamma_{i,i'} = |[\bm{\tau_r}]_i - [\bm{\tau_r}]_{i'}|$, $\zeta_{i,i'} = |[\bm{\tau_c}]_i - [\bm{\tau_c}]_{i'}|$, and $\xi_{i,i'} = |[\bm{\tau_r}]_i - [\bm{\tau_c}]_{i'}|$.  Denote $\Delta_r = \textrm{min}_{i\neq i'} \gamma_{i,i'}$, $\Delta_c = \textrm{min}_{i\neq i'} \zeta_{i,i'}$, $\Delta_{rc} = \textrm{min}_{i,i'} \xi_{i,i'}$, and $\Delta = \textrm{min} (\Delta_r, \Delta_c, \Delta_{rc})$. The following Proposition~\ref{prep:condition} states that the condition number is governed by the number of samples and the joint minimum separation.
\begin{proposition}
  The condition number $\kappa_g$ of matrix $\mathbf{V_g}$ in \eqref{eq:problemjoint} satisfies $\kappa_g^2 \leq (N - 1/\Delta + 1)/(N - 1/\Delta - 1)$.
  \label{prep:condition}
\end{proposition}
\begin{IEEEproof}
    Recall the received signal in \eqref{eq:problemjoint}. %To bound the condition number of $\mathbf{V_g}$,
    Consider the Dirac comb $h(n)= \sum_{t}\delta(n-t)= \sum_{t'} e^{ \textrm{j} 2 \pi t' n}$. The norm of the measurements is
    \begin{align}
       &\sum_{n=0}^{N-1} \| [\mathbf{y}]_n\|^2 = \int h(n) I_{E}(n) \|Y(n)\|^2 \partial n \nonumber   \\   
       & \leq \int h(n) C_{E}(n) \|Y(n)\|^2 \partial n =
       \int h(n) C_{E}(n) Y(n)Y^*(n) \partial n \nonumber \\
       &=  
       \Big(\sum_{i, i' =1}^K[\bm{\beta}]_i [\bm{\beta}]^*_{i'} \int h(n) C_{E}(n) |\widetilde{g}_r(n)|^2 \alpha_{\gamma_{i,i'}}^n   \partial n \nonumber \\
       &+\sum_{i, i' =1}^{K,Q} [\bm{\beta}]_i [\bm{\omega}]^*_{i'}  \int h(n) C_{E}(n) \widetilde{g}_r(n) \widetilde{g}^*_c(n) \alpha_{\xi_{i,i'}}^n   \partial n \\
       &+\sum_{i, i' =1}^{Q, K} [\bm{\omega}]_i [\bm{\beta}]^*_{i'} \int h(n) C_{E}(n) \widetilde{g}^*_r(n) \widetilde{g}_c(n) \alpha_{\xi_{i,i'}}^n   \partial n\nonumber \\
       &+ \sum_{i, i'=1}^Q [\bm{\omega}]_i [\bm{\omega}]^*_{i'} \int h(n) C_{E}(n) |\widetilde{g}_c(n)|^2 \alpha_{\zeta_{i,i'}}^n   \partial n\Big)\nonumber,
    \end{align}  
 where the inequality follows from replacing $I_E(n)$ by $C_E(n)$. Substituting the expression of the Dirac comb $h(n)$,  $g_{r,c}(n)=\widetilde{g}_r(n) \widetilde{g}^*_c(n)$, and $g_{c,r}(n)=\widetilde{g}_c(n) \widetilde{g}^*_r(n)$ produces 
    \begin{align}
       &\sum_{n=0}^{N-1} \| [\mathbf{y}]_n\|^2  %\\
       %&
       \leq \Big( \sum_{t'} \sum_{i, i' =1}^K[\bm{\beta}]_i [\bm{\beta}]^*_{i^{'}}  \mathcal{F} \left\{ C_{E}(n) |\widetilde{g}_r(n)|^2 \alpha_{\gamma_{i,i^{'}}}^n \right\}\nonumber \\
       &+\sum_{t'}\sum_{i, i' =1}^{K,Q} [\bm{\beta}]_i [\bm{\omega}]^*_{i'}  \mathcal{F}\left\{ C_{E}(n) g_{r,c}(n) \alpha_{\xi_{i,i'}}^n \right\} \\
       &+\sum_{t'} \sum_{i, i' =1}^{Q, K} [\bm{\omega}]_i [\bm{\beta}]^*_{i'}  \mathcal{F}\left\{ C_{E}(n) g_{c,r}(n) \alpha_{\xi_{i,i'}}^n \right\}\nonumber \\
       &+\sum_{t'}  \sum_{i, i'=1}^{Q} [\bm{\omega}]_i [\bm{\omega}]^*_{i'}  \mathcal{F}\left\{  C_{E}(n) |\widetilde{g}_c(n)|^2 \alpha_{\zeta_{i,i'}}^n \right\} \Big)\nonumber.
    \end{align}
Using the convolution theorem and $\widetilde{C}_E = \mathcal{F}\{C_E\}$ yields 
    \begin{align}
        & \| [\mathbf{y}]_n\|^2 
       \leq \Big( \sum_{t'} \sum_{i, i' =1}^K[\bm{\beta}]_i [\bm{\beta}]^*_{i^{'}}  \widetilde{C}_{E}(t'+\gamma_{i,i^{'}}) \ast \mathcal{F} \{\widetilde{g}_r(n)^2 \alpha_{\gamma_{i,i^{'}}}^n \} \nonumber \\
       &+\sum_{t'}\sum_{i, i' =1}^{K,Q} [\bm{\beta}]_i [\bm{\omega}]^*_{i'}  \widetilde{C}_{E}(t'+\xi_{i,i'}) \ast \mathcal{F}\left\{g_{r,c}(n) \alpha_{\xi_{i,i'}}^n \right\}
       \end{align}
       \begin{align}
       &+\sum_{t'} \sum_{i, i' =1}^{Q, K} [\bm{\omega}]_i [\bm{\beta}]^*_{i'}  \widetilde{C}_{E}(t'+\xi_{i,i'}) \ast \mathcal{F}\left\{g_{c,r}(n) \alpha_{\xi_{i,i'}}^n \right\} \nonumber\\
       &+\sum_{t'}  \sum_{i, i'=1}^{Q} [\bm{\omega}]_i [\bm{\omega}]^*_{i'}    \widetilde{C}_{E}(t'+\zeta_{i,i'}) \ast \mathcal{F}\left\{|\widetilde{g}_c(n)|^2 \alpha_{\zeta_{i,i'}}^n \right\} \Big).\nonumber\label{eq:yn}
    \end{align}

Note that $\widetilde{C}_E$ is supported on $[-\Delta, \Delta]$. Hence, the first and fourth terms are non-zero only when $t'=0$ and $i=i'$. The second and third terms are zero for all $t', i, i'$ because $|t' +\xi_{i,i'}|> \Delta$, for all $t',i,i'$. Substituting $G_{ri}(n)= |\beta_{i}|^2\mathcal{F} \{ |\widetilde{g}_r(n)|^2 \alpha_{\gamma_{i,i}}^n \}$ , $G_{ci}(n) =|\omega_i|^2 $ $ \mathcal{F}\{|\widetilde{g}_c(n)|^2 \alpha_{\zeta_{i,i}}^n \}$ in \eqref{eq:yn} yields 
    \begin{align}
       & \| [\mathbf{y}]_n\|^2 \leq  
        \sum_{i}^K  \widetilde{C}_{E}(0) \ast G_{ri}(n)+ \sum_{i}^{Q} \widetilde{C}_{E}(0) \ast G_{ci}(n).
    \end{align}
From \cite[Theorem 2.2]{moitra2015super}, the integral of the majorant function, %$C_E(t)$ is equal to $2n+1/\Delta$, 
i.e., $\int C_E(t) = 2n + 1/\Delta$. Hence, %-[theorem 2.2] 
we obtain 
    \begin{align}             &\Big\|\mathbf{V_g}\bm{\psi} \Big\|^2=\sum_{n=0}^{N-1} \| [\mathbf{y}]_n\|^2 \leq (2n + \frac{1}{\Delta}) \times G(n),
    \end{align}
where $G(n)= \sum_{i}^K  G_{ri}(n)+ \sum_{i}^{Q} G_{ci}(n)$. The lower bound is similarly obtained from the minorant $c_E(n)$, %defined in \cite{moitra2015super} 
whose Fourier transform is supported on $[-\Delta, \Delta]$, as
    \begin{align}
         &\Big\|\mathbf{V_g}\bm{\psi} \Big\|^2=\sum_{n=0}^{N-1} \| [\mathbf{y}]_n\|^2 \geq (2n - \frac{1}{\Delta})\times  G(n),
    \end{align}
Defining $\kappa_g^2 = \big(\text{max}_{\bm{\nu}}\|\mathbf{V_g \bm{\nu}}\|^2/\text{min}_{\bm{\nu}}\|\mathbf{V_g \bm{\nu}}\|^2\big)$ and $N=2n+1$ results in $\kappa_g^2 \leq (N+\frac{1}{\Delta} - 1)/(N-\frac{1}{\Delta} - 1)$. 
\end{IEEEproof}

From Proposition \ref{prep:condition}, the condition number depends on the minimum joint separation. This gives $N>(1/\Delta + 1)$, this shows the coupled nature of radar and communications in the DBD problem. %Note that it is related to the condition number presented by Moitra, A.  \cite{moitra2015super}. However, this work extends it to a blind case in a co-existence radar communications scenario since $\Delta$ depends on both radar and comms.
\vspace{-8pt}
\section{Low-Rank Hankel Matrix Recovery}
The recovery of $\bm{\tau}_r$, $\bm{\tau}_c$, $\bm{\beta}$, and $\bm{\omega}$ poses challenges due to the ill-posed nature of the problem. To address this, we reasonably assume that the unknown point spread functions (PSF), $\mathbf{g_r}$ and $\mathbf{g_c}$, can be accurately represented within a known low-dimensional subspace. This assumption is justified by considering the rank of the Vandermonde matrices, which is influenced by the number of radar and communications delays which is usually smaller than the number of measurements \cite{chen2022vectorized, vargas2022joint, vargas2023dual, vouras2022overview}. Following this, we denote $\mathbf{\widetilde{g}_r} = \mathbf{Bh_r}$ and $\mathbf{\widetilde{g}_c} = \mathbf{Dh_c}$, %as
%\begin{equation}
%     \begin{aligned}
%    \widetilde{\mathbf{g_r}} = \mathbf{Bh_r}, \     \widetilde{\mathbf{g_c}} = \mathbf{Dh_c},
%    \end{aligned}
%\end{equation}
where $\mathbf{B}\in \mathbb{C}^{N\times N_r}$, $\mathbf{h_r}\in \mathbb{C}^{N_r}$, $\mathbf{D}\in \mathbb{C}^{N\times N_c}$, and $\mathbf{h_c}\in \mathbb{C}^{N_c}$. Collect the unknown radar and communications variables, respectively, in the matrices %$\mathbf{X}_r \in \mathbb{C}^{N_r\times N}$ and $\mathbf{X}_c \in \mathbb{C}^{N_c\times N}$ as
%\begin{align}
    $\mathbf{X_r} = \sum_{i=1}^K [\bm{\beta}]_i \mathbf{h_r} \bm{a}_{[\bm{\tau_r}]_i}^T\;\in \mathbb{C}^{N_r\times N}$ and %,\nonumber\\ 
    $\mathbf{X}_c = \sum_{l=1}^Q [\bm{\omega}]_i \mathbf{h_c} \bm{a}_{[\bm{\tau_c}]_l}^T\;\in \mathbb{C}^{N_c\times N}$,
    %\label{eq:XrXc}
%\end{align}
with $\bm{a}_{[\bm{\tau}_r]_i}^T = [1, e^{-\textrm{j}2\pi [\bm{\tau}_r]_i}, e^{-\textrm{j}2 \pi [\bm{\tau}_r]_i (2)}, \ldots, e^{-\textrm{j}2\pi[\bm{\tau}_r]_i (N-1)}]$ is the vector containing all the atoms $\alpha_{[\bm{\tau}_r]_{i}}^n$; $\bm{a}_{[\bm{\tau}_c]_l}$ is defined similarly using atoms $\alpha_{[\bm{\tau}_c]_{l}}^n$. Denote $\mathbf{s}_j=[\mathbf{b}_j^T ,\mathbf{d}_j^T]^T$, where $\mathbf{b}_j$  and $\mathbf{d}_j$ are the $j$-th columns of $\mathbf{B}^*$ and $\mathbf{D}^*$, respectively. Rewrite \eqref{eq:problem1} as
\begin{align}
    [\mathbf{y}]_j=\langle \mathbf{s}_j\mathbf{e}_j^T, \mathbf{X}\rangle,
    \label{eq:innerproduct}
\end{align}
with $\mathbf{X}=[\mathbf{X_r}^T,\mathbf{X_c}^T]^T \in \mathbb{C}^{(N_r+N_c)\times N}$, and $\mathbf{e}_j\in \mathbb{R}^{N}$ is the $j$-th canonical vector. %Without loss of generality, 
For the sake of simplicity, consider $N_c=N_r = N_{rc}$. Define a linear operator $\mathcal{A}: \mathbb{C}^{2N_{rc} \times N} \xrightarrow[]{}\mathbb{C}^N$ such that $[\mathcal{A}(\mathbf{X})]_j=\langle \mathbf{s}_j\mathbf{e}_j^T, \mathbf{X}\rangle$. Then, \eqref{eq:innerproduct} becomes
\begin{equation}
    \mathbf{y}=\mathcal{A}(\mathbf{X}).
\end{equation}
Select %two number 
$N_1, N_2 \in \mathbb{N}$ such that $N_1+N_2=N+1$. The low-dimensional subspace assumption allows us to represent the unknown variables in a low-rank matrix. To construct such a matrix, apply a linear operator $\mathcal{H(\cdot)}$ to both $\mathbf{X_r}$ and $\mathbf{X_c}$ such that it produces a Hankel matrix of higher dimensions using the columns of the input matrix. Then, concatenating both Hankel matrices using the operator  $\mathcal{C}(\mathbf{X}) =  \mathcal{C}([\mathbf{X_r}^T,\mathbf{X_c}^T]^T) = [\mathcal{H(\mathbf{X_r})}, \mathcal{H(\mathbf{X_c})}] $ yields  %construct the following \textcolor{blue}{variation of the vectorized Hankel matrix} associated with $\mathbf{X}$ as %in \cite{chen2022vectorized} given by
\begin{equation}
    \mathcal{C}(\mathbf{X})=\begin{bsmallmatrix}
       \mathbf{x_r}_{0} & \ldots &  \mathbf{x_r}_{N_2-1} & \mathbf{x_c}_{0}   & \ldots & \mathbf{x_c}_{N_{2}-1}\\
       \mathbf{x_r}_{1} & \ldots &  \mathbf{x_r}_{N_2} & \mathbf{x_c}_{1}   & \ldots & \mathbf{x_c}_{N_{2}}\\
       \vdots & \vdots & \vdots & \vdots & \ddots & \vdots \\
        \mathbf{x_r}_{N_1-1} & \ldots &  \mathbf{x_r}_{N-1} & \mathbf{x_c}_{N_1-1}   & \ldots & \mathbf{x_c}_{N-1}\\
    \end{bsmallmatrix}. %\in \mathbb{R}^{N_1 N_{rc} \times 2N_2}
\end{equation}
The concatenated matrix is decomposed as
\begin{equation}
    \mathcal{C}(\mathbf{X}) = \mathbf{V_{h,\alpha}} \text{diag}(\bm{\psi}) \mathbf{V}_{\alpha}^T,
    \label{eq:vdv}
\end{equation}
where the $(N_1N_{rc})\times (K+Q)$ complex matrix $\mathbf{V_{h,\alpha}}$ is
    \begin{align}
        \mathbf{V_{h,\alpha}}&=
            \begin{bsmallmatrix}
                \mathbf{h_r} & \ldots & \mathbf{h_r} & \mathbf{h_c} & \ldots & \mathbf{h_c}\\
                \mathbf{h_r}\alpha_{[\bm{\tau}_r]_1}^1 & \ldots & \mathbf{h_r} \alpha_{[\bm{\tau}_r]_{K}}^1 & \mathbf{h_c}\alpha_{[\bm{\tau}_c]_1}^1 & \ldots & \mathbf{h_c}\alpha_{[\bm{\tau}_c]_{Q}}^1\\
                \vdots & \ddots & \vdots & \vdots  & \ddots & \vdots  \\
                \mathbf{h_r}\alpha_{[\bm{\tau}_r]_1}^{N_1-1} & \ldots & \mathbf{h_r}\alpha_{[\bm{\tau}_r]_{K}}^{N_1-1} & \mathbf{h_c}\alpha_{[\bm{\tau}_c]_1}^{N_1-1} & \ldots & \mathbf{h_c}\alpha_{[\bm{\tau}_c]_{Q}}^{N_1-1}
            \end{bsmallmatrix},%\in \mathbb{C}^{(N_1N_{rc})\times (K+Q)},
            \label{eq:Vh}
    \end{align}
    with $\alpha_{[\bm{\tau}_r]_{i}}^{n_1}$ and $\alpha_{[\bm{\tau}_c]_{l}}^{n_1}$ are as defined in \eqref{eq:problemjoint} except that $n_1=0,\cdots,N_1-1$ and $n_2=0,\cdots,N_2-1$, and %the matrix $\mathbf{V}^T\in \mathbb{R}^{(K+Q)\times 2\cdot n_2}$ is
\begin{align}
        \mathbf{V}_{\alpha}^T=\begin{bsmallmatrix}
    \bm{\alpha}_{r}^0 & .. & \bm{\alpha}_{r}^{N_2-1} & \mathbf{0} & .. & \mathbf{0} \\
    \mathbf{0} & .. & \mathbf{0} & \bm{\alpha}_{c}^0 & .. & \bm{\alpha}_{c}^{N_2-1} 
    \end{bsmallmatrix},\;\in \mathbb{R}^{(K+Q)\times 2N_2},
    \label{eq:vt}
\end{align}
with $\bm{\alpha}_{r}^{n_2} = [e^{-\textrm{j}2\pi {n_2}[\bm{\tau}_r]_1 }, e^{-\textrm{j}2\pi {n_2} [\bm{\tau}_r]_2}, \ldots, e^{-\textrm{j}2\pi {n_2} [\bm{\tau}_r]_K }]^T \in \mathbb{R}^{K}$ and $\bm{\alpha}_{c}^{n_2} = [e^{-\textrm{j}2\pi {n_2} [\bm{\tau}_c]_1 }, e^{-\textrm{j}2\pi {n_2} [\bm{\tau}_c]_2}, \ldots, e^{-\textrm{j}2\pi {n_2} [\bm{\tau}_c]_Q }]^T \in \mathbb{R}^{Q}$. 

From \eqref{eq:vdv}, $\text{rank}(\mathcal{C}(\mathbf{X})) \le (K+Q)$. Hence, $\mathcal{C}(\mathbf{X})$ is a low-rank structure. Then, the problem of recovering $\mathbf{X}$ becomes
\begin{align}
  \underset{\mathbf{X} \in \mathbb{R}^{2 \cdot N_{rc}\times N}}{\text{ argmin}}
   \|\mathcal{C}(\mathbf{X})\|_*\text{ subject to } \mathbf{y}=\mathcal{A}(\mathbf{X}).
  \label{eq:opt3}
\end{align}
%Assume the optimal solution of \eqref{eq:opt3} is $\mathbf{X}^{\star}$. 
An optimal solution to \eqref{eq:opt3} is guaranteed \cite{chen2022vectorized} if $\sigma_{\textrm{min}}(\mathbf{V_{h,\alpha}}\mathbf{V_{h,\alpha}}^T)\geq N_1/\mu$ and $ \sigma_{\textrm{min}}(\mathbf{V}_{\alpha}\mathbf{V}_{\alpha}^T)\geq N_2/\mu$ for $\mu > 1$. From Proposition \ref{prep:condition} and \cite{chen2022vectorized}, these conditions imply $\Delta > 2\mu / N(\mu-1)$. Once the matrix $\mathbf{X}$ is obtained, the delays $\bm{\tau_r},\bm{\tau_c}$ are recovered via MUSIC. % algorithm proposed in \cite{chen2022vectorized}. 
Finally, the waveforms $\mathbf{g_r},\mathbf{g_c}$ are retrieved using the least-squares method \cite{vargas2023dual}.

%\begin{assumption}
%There exist a constant $\mu>1$ such that
%\begin{align}
%    &\sigma_{\textrm{min}}(\mathbf{V}_h \mathbf{V}_h^T)>\frac{n_1}{\mu}\  \text{ and }
%    & \sigma_{\textrm{min}}(\mathbf{V} \mathbf{V}^T)>\frac{n_2}{\mu}
%\end{align}
%\end{assumption}

%\begin{proposition}
%if $\Delta > 2\mu/(n(\mu-1)) $ then $\mathbf{X}$ is the only solution to $\mathcal{H}(X)$
%\end{proposition}
\vspace{-8pt}
\section{Numerical Experiments}
We compared our approach to DBD against SoMAN SDP in \cite{vargas2023dual}, which assumes a low-dimensional subspace structure of the waveform $\mathbf{g_r}$ and message $\mathbf{g_c}$ and additionally exploits the sparsity of the channels. The SoMAN method represents the matrices $\mathbf{X_r} $ and $\mathbf{X_c}$ %in \eqref{eq:XrXc} 
as a linear combination of atoms given by the sets 
%\begin{align}
    $\mathcal{A}_r = \Big\{\mathbf{h_r}\bm{a}_{\tau_r}^H: \tau_r \in [0,1) \Big\}$ and %\; , 
    $\mathcal{A}_c = \Big\{\mathbf{h_c}\bm{a}_{\tau_c}^H, \tau_c \in [0,1) \Big\} \;\subset \mathbb{C}^{N_{rc}\times N}$. 
%    \label{eq:atomic_set_com}
%\end{align}
 This leads to the following formulation of \textit{atomic norms} %$||\mathbf{X}_r||_{\mathcal{A}_r}$ and $||\mathbf{X}_c||_{\mathcal{A}_c}$ %- a sparsity-enforcing analog of $\ell_1$ norm 
 %, respectively, general atomic sets $\mathcal{A}_r$ and $\mathcal{A}_c$:
\begin{align}
    &||\mathbf{X}_r||_{{\mathcal{A}_r}} = \inf \Bigg\{\sum_\ell^{K} |[\bm{\beta}]_{l}| \Big| {\mathbf{X}}_r = \sum_\ell [\bm{\beta}]_{l} \mathbf{h_r}\bm{a}_{[\bm{\tau_r}]_i}^H\Bigg\},
    \label{eq:atomic_norm_rad} \\
    &||\mathbf{X}_c||_{\mathcal{A}_c} = \inf \Bigg\{\sum_q^Q |[\bm{\omega}]_{q}| \Big| \mathbf{X}_c = \sum_q [\bm{\omega}]_{q} \mathbf{h_c}\bm{a}_{[\bm{\tau_c}]_i}^H\Bigg\}.
    \label{eq:atomic_norm_com}
\end{align}
The SoMAN minimization estimates $\mathbf{X}_r$ and $\mathbf{X}_c$ as %of $||{\mathbf{X}_r}||_{\mathcal{A}_r} + ||{\mathbf{X}_c}||_{\mathcal{A}_c}$ given by
\begin{align}
    &\min_{{\mathbf{X}}_r,{\mathbf{X}}_c} ||{\mathbf{X}}_r||_{\mathcal{A}_r} +||{\mathbf{X}}_c||_{\mathcal{A}_c} %\nonumber\\
    %&
    \;\text{s. t.}\;     \mathbf{y} = \aleph_r({{\mathbf{X}}_r}) + \aleph_c({\mathbf{{X}}_c}).
    \label{eq:primal_problem}
\end{align}
where $[\aleph_r({{\mathbf{X}}_r})]_i = \textrm{Tr}(\mathbf{e}_i\mathbf{h_r}^H\mathbf{X}_r)$ and  $[\aleph_c({{\mathbf{X}}_c})]_j =  \textrm{Tr}(\mathbf{e}_j\mathbf{h_c}^H\mathbf{X}_c)$. 

%------------------------------------------------------------
\begin{figure}[t]
    \centering
    \includegraphics[width=0.48\textwidth]{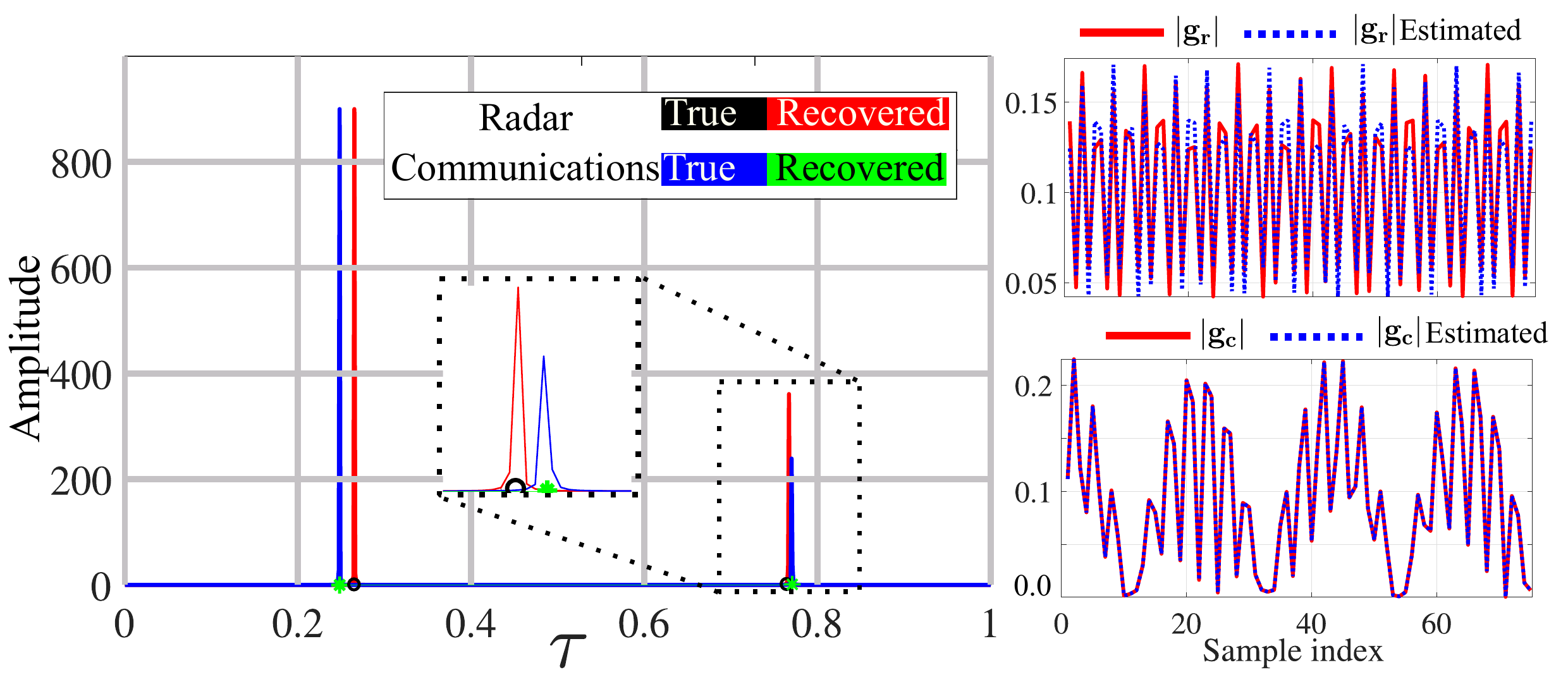}
    \caption{\scriptsize{The pseudo-spectrum computed using the MUSIC algorithm and the true radar $\bm{\tau_r}$ and communications $\bm{\tau_c}$ delays (left), recovered radar waveform (right top), and communications message (right bottom) with $N=75$ samples, $K=2$ targets and $Q=2$ paths and subspace dimensions $N_r=N_c =2$. The NMSE obtained is $0.09987$ and $2.3144e-5$ for the radar waveform $\mathbf{g_r}$ (top-right) and communications message $\mathbf{g_c}$ (bottom-right), respectively. }
    }
    \label{fig:my_label}
\end{figure}
%------------------------------------------------------------
%------------------------------------------------------------

%------------------------------------------------------------
To evaluate the performance of the proposed vectorized Hankel method, we performed several computational simulations using CVX \cite{cvx, gb08} library in MATLAB with SDPT3 \cite{sdpt3} solver. %First, we obtained the phase transition of the proposed method and compared it against SoMAN minimization. 
\begin{figure}[ht]
    \centering
    \includegraphics[width=0.48\textwidth]{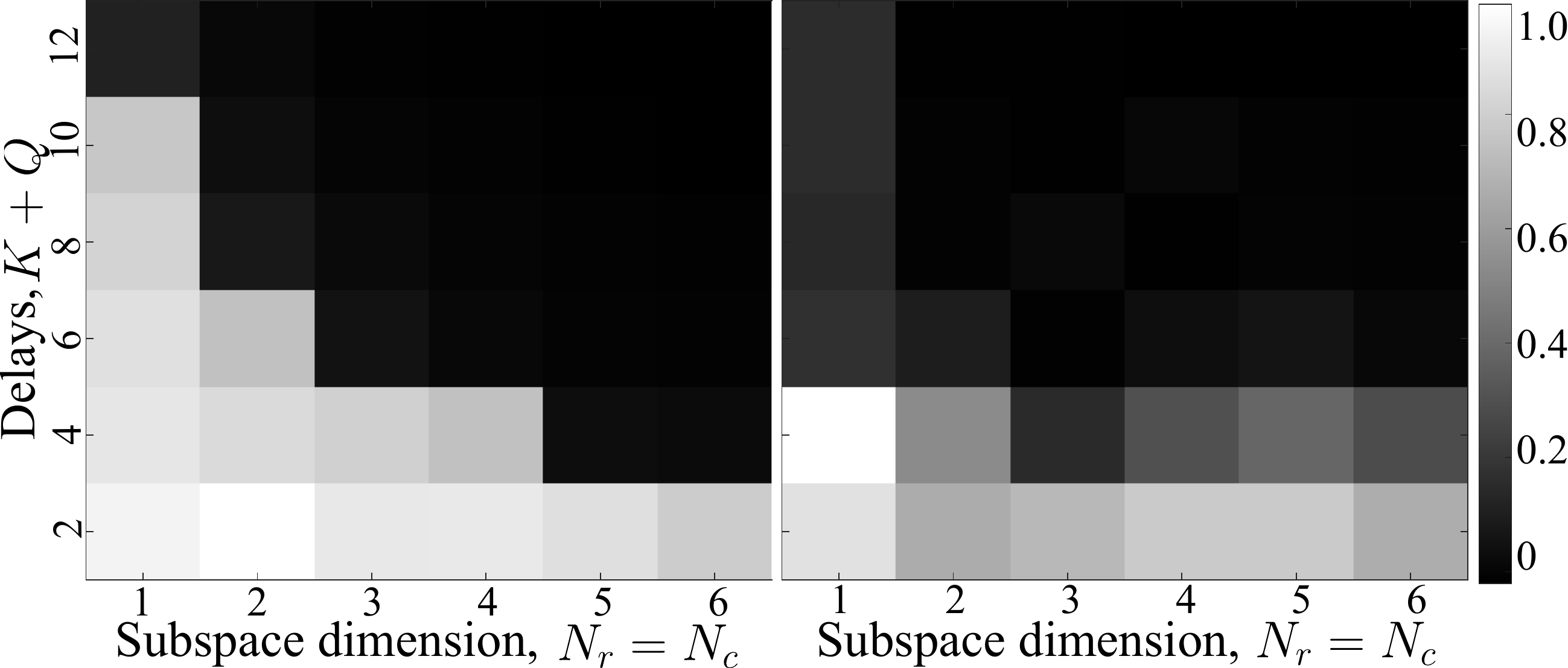}
    \caption{\scriptsize{The probability of reconstruction, averaged over 20 realizations with $N=75$, for the proposed modified Hankel matrix recovery (left) and SoMAN SDP (right). } \vspace{-16pt} }
    \label{fig:phase}
\end{figure}
The delays $\bm{\tau_r}$ and $\bm{\tau_c}$ were drawn uniformly at random from the interval $[0,1)$; the amplitudes $\bm{\beta}, \bm{\omega}$ were generated following $(1+ 10^{\gamma})e^{-j\phi}$ with $\gamma$ sampled from a uniform distribution $[0,1)$ and $\phi$ from $[0,2\pi)$. The columns of the matrices $\mathbf{B}$ and $\mathbf{D}$ are random columns of the DFT matrix of the corresponding size. 

Figure \ref{fig:my_label} shows a specific delay recovery with two radar targets, two communication paths, and two unknown PSFs with $N_r=N_c=2$, $K=Q=2$, and 75 samples. The pseudospectrum is computed using MUSIC \cite{chen2022vectorized}. The proposed method is able to recover the exact position of the delays. Additionally, the recovered radar waveform and communications message are accurately estimated with the normalized mean-squared error (NMSE) $=\|\mathbf{g}-\mathbf{g}^*\|/\|\mathbf{g}\|$ of $0.0998$ and $2.31\times 10^{-5}$ respectively. %After $\bm{\tau_r}$ and $\bm{\tau_c}$ are recovered, the radar and communications waveforms are estimated using the least squares method \cite{vargas2023dual}.  
Figure \ref{fig:phase} shows the reconstruction probability of the matrices $\mathbf{X}_r$ and $\mathbf{X}_c$ over 20 experiments, where success is declared if the NMSE between the truth and recovered $\mathbf{X}$ is lower than $10^{-3}$. We observe that the phase transition curve of the proposed method is higher than the SoMAN approach. %This phase transition indicates that our method outperform the ANM method.
\vspace{-10pt}
\section{Summary}
We provided an optimal separation condition for the DBD problem using extremization functions. This condition depends on the separation of the radar and communications delays and, contrary to prior guarantees, is jointly imposed. Our recovery algorithm based on a modified vectorized Hankel matrix recovery outperforms SoMAN SDP. Future investigations require generalizing the proposed approach to other coexistence models considered in \cite{vargas2023dual}.
\vspace{-8pt}
%Beurling-Selberg majorization and minorization problem seeks for entire functions $F: \mathbb{C} \xrightarrow{} \mathbb{C}$ and $T: \mathbb{C} \xrightarrow{} \mathbb{C}$ of type at least $2\pi \delta$ which minimizes 
%\begin{align}
%    \int F(x)-f(x) \partial x \\
%    \int f(x) - T(x) \partial x
%\end{align}
%where $f(x)$ is  a function that satisfies that $T(x)\leq f(x) \leq F(x)$\cite{10.2307/23513054, moitra2015super}. A particular case is given when $f(x)=I_{E}(x)$, where $I_{E}(x)$ is the indicator function for a interval $E=[0,N-1]$. 

%\clearpage
\bibliographystyle{IEEEtran}%IEEEtran
\bibliography{main}

\end{document}